\documentclass[prl,amsmath,amssymb,showpacs,showkeys,twocolumn]{revtex4}
\usepackage{graphicx}
\usepackage{dcolumn}
\usepackage{bm}
\usepackage{amssymb}
\usepackage{amsmath}
\begin{document}

\title{Observation of Coherent Precession of Magnetization in Superfluid $^3$He A-phase}

\author{T.~Sato$^1$}
\author{T.~Kunimatsu $^2$}
\author{K.~Izumina$^1$}
\author{A.~Matsubara$^{1,2,3}$}
\author{M.~Kubota$^1$}
\author{T.~Mizusaki$^{2,3}$}
\author{Yu.~M.~Bunkov$^{1,4}$\/\thanks{e-mail: yuriy.bunkov@grenoble.cnrs.fr}}

\affiliation{
$^1$Institute for Solid State Physics, The University of Tokyo, Chiba 277-8581, Japan \\
$^2$Department of Physics, Graduate School of Science, Kyoto University, Kyoto 606-8502, Japan \\
$^3$Research Center for Low Temperature and Materials Sciences, Kyoto University, Kyoto 606-8502, Japan \\
$^4$MCBT, Institut N\'{e}el, CNRS/UJF, 38042, Cedex9, France}
\date{\today}


\begin{abstract}

We report the first observation of coherent quantum precession of magnetization (CQP) in superfluid $^3$He-A in aerogel.  The coherent precession in bulk $^3$He A-phase is unstable due to the positive feedback of spin supercurrent to the gradient of phase of precession.  It was predicted that the homogeneous precession will be stable if the orbital momentum of $^3$He-A could be oriented along the magnetic field.  We have succeeded to prepare this configuration by emerging $^3$He in uniaxially-deformed anisotropic aerogel.  The dissipation rate of coherent precession states in aerogel is much larger then one in  bulk $^3$He-B.  We propose a mechanism of this dissipation.

\end{abstract}

\pacs{67.30.he, 05.30.Jp, 67.30.hj}

\keywords{superfluid $^3$He, spin supercurrent, NMR}

\maketitle



The coherent precession of magnetization was observed early in
superfluid $^3$He-B. Due to the concave shape of dipole-dipole
interaction and negative feed back of spin supercurrent (the quantum
transport of magnetization due to the gradient of the phase of spin
part of order parameter), the coherent precession of magnetization
in $^3$He-B arises spontaneously even in inhomogeneous magnetic
field, as was discovered in 1984 by Borovik-Romanov et
al.\cite{HPD}. This effect was named a Homogeneously Precessing
Domain (HPD), due to the splitting of the magnetization in the cell
into two domains, one stationary and one with the coherent
precession of magnetization deflected on the angle slightly above
the magic angle of $104^\circ$.  Recently the HPD was identified as
magnon Bose-Einstein condensation by Bunkov and Volovik~\cite{BV2}.

Here we report the first observation of the coherent quantum
precession of magnetization (CQP) in A-like phase in uniaxially
deformed aerogel, where the $\hat{l}$-vector in the orbital part of
the order parameter is oriented along the magnetic field.  We call
the coherent precession in A-phase as the CQP, distinguishing it
from the HPD in B-phase and will discuss later a basic difference in
both modes.

In bulk $^3$He A-phase, the homogeneous precession is unstable even
in homogeneous magnetic field because of the convex shape of dipole
energy potential~\cite{InstabAF,InstabAB}.  The dipole interaction
in A-phase depends on the orientations of the order parameter
denoted by two vectors, the orbital part $\hat{l}$ and the spin part
$\hat{d}$ of the order parameter.  We consider the uniform motion of
magnetization in $^3$He A-phase in high magnetic fields as,
\begin{equation}
M_\perp=|M| \sin\beta \sin(\omega t + \varphi_0), \label{eq:1}
\end{equation}
where the transverse magnetization $M_\perp$ rotates with a constant
tipping angle $\beta$ from the magnetic field $\vec H$ and an
angular frequency $\omega$, $|M|=\chi_A H$ and $\chi_A$ is the
susceptibility of the A-phase. The dipole energy $V_D(\beta,
\lambda)$ averaged over the fast precession of $M_\perp$ is
calculated as a function of $\beta$ for various values of the angle
$\lambda$ between $\hat{l}$ and $\vec H$~\cite{BV}. The NMR
frequency $f=\omega/2\pi$ in Eq.~(\ref{eq:1}) under the dipole
interaction $V_D(\beta, \lambda)$ with a finite tipping angle
$\beta$ is obtained by
\begin{equation}
\omega = \omega_L-\partial V_D/\partial (|M|\cos\beta),
\end{equation}
where the Larmor angular frequency $\omega_L=\gamma H$. The
stability condition of the uniform motion of magnetization of
Eq.~(\ref{eq:1}) is written as $\partial^2 V_D/\partial
(|M|\cos\beta)^2 > 0$. In the opposite case the uniform motion is
unstable and decays into the non-uniform motion of
magnetization~\cite{InstabAF,InstabAB}.

\begin{figure}[ttt]
 \includegraphics[width=0.4\textwidth]{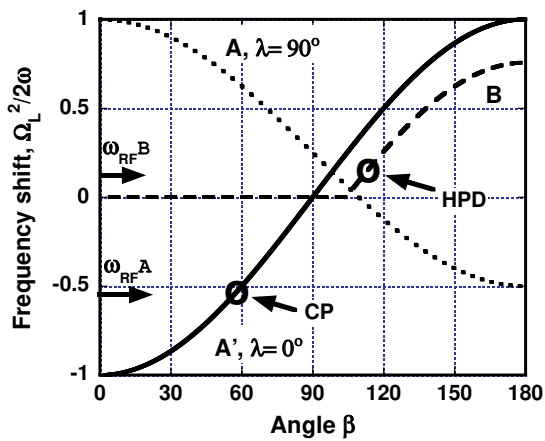}
 \caption{The frequency shift from Larmor frequency
 versus the tipping angle $\beta$; solid curve for $\lambda = 0^\circ$ and
dotted curve for $\lambda = 90^\circ$ for A-phase and dashed curve
for B-phase. When a sufficiently large rf-field is applied, the CQP
for $\lambda = 0^\circ$ is excited at a finite $\beta$ for a whole
sample and the HPD is excited at $\beta \sim 104^\circ$ for the
precessing domain. } \label{schema}
\end{figure}

In Fig.~\ref{schema}, the NMR frequency shifts, $\Delta\omega =
\omega-\omega_L = - \partial V_D/\partial (|M|\cos\beta)$, are shown
as functions of the tipping angle $\beta$ for both cases of $\lambda
= 90^\circ$ (dotted curve labeled by A) and $0^\circ$ (solid curve
labeled by A'). The orientation of the $\hat{l}$-vector in bulk
A-phase is determined by minimizing the dipole energy and the
$\hat{l}$-vector is oriented perpendicularly to the $\vec H$.  Since
the direction of the $\hat{l}$-vector is fixed during cw- and pulsed
NMR, the CQP is always unstable in bulk A phase, which has been
confirmed experimentally.  On the other hand, the CQP is stable in
the case that the $\hat{l}$-vector is parallel to the field, which
is realized when A-like phase is immersed in aerogel squeezed along
the magnetic field direction~\cite{KunimatsuJETPLett}. The stability
condition for the HPD in B-phase is essentially the same as the
above argument.

The NMR frequency shift in B-phase is shown in Fig.~\ref{schema} by
the broken curve labeled by B and the frequency shift from Larmor
frequency appears only for angles $\beta>104^\circ$.  Therefore, the
uniform motion is stable only for $\beta > 104^\circ$.  If there is
a small field gradient, the deflected magnetization is redistributed
into two domains in the cell in such a way that the region for lower
magnetic fields forms the one domain with angles of deflection more
then $104^\circ$ and the region for higher magnetic fields forms the
other domain with $\beta=0$.  This is why the coherent precession in
B-phase is called as the HPD.  On the contrary, the CQP in the
branch of A' in A-phase can be stabilized even at small angles and
does not split into two domains. It should be noted that the CQP and
HPD are self-organized states of macroscopic coherent precession
even under inhomogeneous external fields, in which the spin
supercurrent flows and the tipping angle is adjusted in such a way
that the gradient of the phase of the precessing magnetization is
automatically canceled.

To orient $\hat{l}$ we used the $^3$He A-phase confined in
uniaxially-deformed aerogel with 98 \% porosity~\cite{Mulders}.
Aerogel plays the role of impurities with randomly distributed
anisotropy
which suppresses the orientational long-range order of $\hat{l}$ and
forms Larkin-Imry-Ma state (LIM)~\cite{LIM}.  However, it was
proposed that when the aerogel sample is globally deformed, and
impurity scattering is not isotropic, the global anisotropy in
scattering length suppressed the LIM state and the long-range order
of $\hat{l}$ is restored~\cite{VolovikAerogel}. We investigated the
A-like phase in uniaxially-deformed aerogel and found that the main
cw-NMR spectrum in A-like phase showed a full negative
shift~\cite{KunimatsuJETPLett}.  We investigated the change of
cw-NMR spectrum under rotation, and studied the global orientation
effect due to anisotropic deformation of aerogel against the flow
orientation effect~\cite{KunimatsuQFS2007} in both A-like and B-like
phases.  A uniaxial deformation of about 2\% along the magnetic
field appears to be sufficient to orient the orbital momentum
$\hat{l}$ along ${\vec H}$. In this letter, all data are taken from
the sample noted by the S-D sample in reference
\cite{KunimatsuJETPLett} at a pressure 29.3 bar in a magnetic field
of 290 gauss, corresponding to an NMR frequency of 940 kHz.  The
sample had the form of a cylinder (the diameter is 5 mm, the length
is 3 mm) with the global anisotropy axis oriented along the external
magnetic field.
We added about 1\% $^4$He to $^3$He sample in order to eliminate
$^3$He solid on aerogel strands. The experiments were performed for
an excitation voltage $v_{rf} =3$ volts at two different
temperatures, $T=0.8 T_{ca}$ and $T=0.7 T_{ca}$, corresponding to
A-like and B-like state, respectively, where the excitation
rf-current $i_{rf}$ is fed through 300 k$\Omega$ resistance. The
superfluid transition temperature in this aerogel is $T_{ca}$=2.07
mK. Figure 2 shows typical data of $M_\perp$({\rm a.u.}) vs.\
$\Delta f$ for the CQP in A-like phase (labeled by A' ) and for the
HPD in B-like phase (labeled by B); solid curves for the
frequency-sweep upward and dotted ones for the sweep downward, where
$M_\perp({\rm a.u.}) = \sqrt{V_{disp}^2 + V_{abs}^2}$, $V_{disp}$ is
the dispersion signal and $V_{abs}$ is the absorption signal. We
actually sweep magnetic fields for a fixed NMR frequency. To observe
the HPD signal a small gradient of magnetic field 2.8 ${\mu}$T/mm
was applied.  During the upward sweep, the HPD starts to form at
zero frequency shift, while the CQP in $^3$He A-phase starts at
negative frequency shift in agreement with Fig.~\ref{schema}.

\begin{figure}[ttt]
  \includegraphics[width=0.4\textwidth]{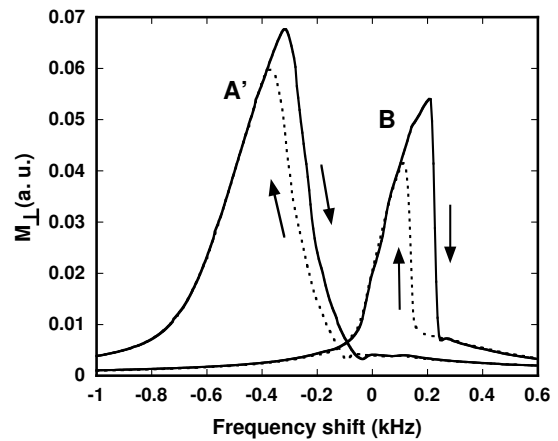}
 \caption{Formation of the coherent precession of magnetization;
the CQP in A-like phase at $0.8 T_{ca}$ (labeled by A') and the HPD
in B phase at $0.7 T_{ca}$ (labeled by B). The solid curve are taken
at an rf-field (3 volts) for upward sweep and the dotted one for
downward sweep. } \label{ABcompere}
\end{figure}

\begin{figure}[ttt]
 \includegraphics[width=0.4\textwidth]{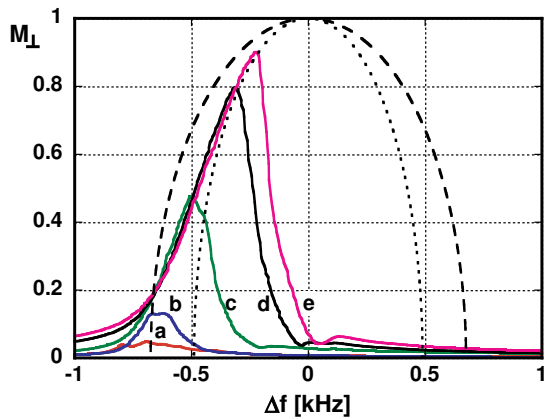}
 \caption{The normalized amplitude of NMR signal by $M_\perp$($\beta=90^\circ$) while sweeping frequency upward at different excitations $v_{rf}$ (a : 0.1 V (only this signal is multiplied by 5 to be visible), b : 0.5 V, c : 1.5 V, d : 3 V, and e : 4 V).
The dashed line corresponds to the theoretical dependence of
$M_\perp$ on the tipping-angle dependent frequency shift given by
$\Delta \omega = - (\Omega_A^2/2\omega L) \cos\beta$ with the
maximum frequency shift chosen at the peak of signal a and the
dotted line corresponds to that for the maximum frequency shift is
chosen at the right edge of signal a.
}
 \label{Amplsmall}
\end{figure}

Figure~\ref{Amplsmall} shows signals from the CQP in $^3$He A-like
phase at different amplitudes of rf-fields, taken for the upward
sweep.  The signals are proportional to the total transverse
magnetization, $M_\perp=\int d^3r \chi_A H \sin\beta$, and we
normalized them to the maximum of the signal which should occur if
the magnetization is deflected by $\beta=90^\circ$ in the whole
sample; i.e. $M_\perp({\rm max})=\chi_AH V$, where $V$ is the volume
of the sample.  Experimentally we cannot reach $M_\perp({\rm max})$
but we can extract it from the measured value of the maximal HPD
signal in $^3$He B-like phase, which corresponds to the
magnetization deflected by $\beta=104^\circ$ and precessing
homogeneously in the whole sample: $M_\perp({\rm HPD}) =\chi_B
HV\sqrt{15/16}$.  Using the known values of magnetic susceptibility
in the two phases, we derive the maximum of the $^3$He A-like phase
signal.

Normalized signals $M_\perp$ at different excitations follow a
universal curve as a function of frequency shift of NMR $\Delta f$,
which corresponds to the tipping angle $\beta$ determined simply by
$\Delta f$ and {\it not} by the amplitude of the rf-fields. Dashed
and dotted lines show the theoretical dependence of $M_\perp$ on the
tipping-angle dependent frequency shift given by $\Delta \omega = -
(\Omega_A^2/2\omega L) \cos\beta$ for two choices of the maximum
frequency shifts, $\Omega_A$. Deviations can be certainly related to
the residual inhomogeneity of the $\hat{l}$-vector in the sample,
which generates the non-uniform frequency shift.

\begin{figure}[ttt]
 \includegraphics[width=0.4\textwidth]{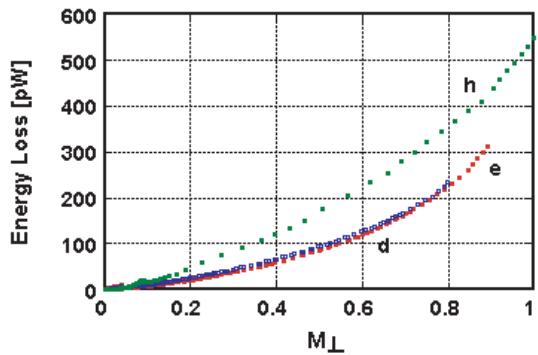}
 \caption{
The energy loss in the CQP as a function of normalized $M_\perp$.
Data denoted by d are for $v_{rf}=3$ V and data denoted by e for
$v_{rf}=4$ V. Loss in the HPD is denoted by h for $v_{rf}=3$ V. }
\label{absorption}
\end{figure}

The energy losses of the CQP and HPD were obtained from the
absorption signal $V_{abs}$ multiplied by the rf current ($= v_{rf}
/ 300$ k$\Omega$). Figure 4 shows the dissipation of the CQP against
$M_\perp$ for two typical excitation levels of rf-fields. The
dissipation for two rf-fields (curves labeled by d for $v_{rf}=3$ V
and e for $v_{rf}=4$ V) falls into a universal curve of the
dissipation, which is proportional to square of $M_\perp$ and does
not depend on $v_{rf}$. Therefore, the dissipation is an intrinsic
property of the CQP in aerogel.  We also show the dissipation of the
HPD vs.\ $M_\perp$ for $v_{rf}=3$ V under a field gradient of 2.8
mT/m by different symbols labeled by b, where $M_\perp$ is the
transverse magnetization normarized by $M_\perp^{\rm max}$ for the
HPD.  The dissipation for the CQP seems to be very large, does not
depend on field gradients for a certain range of the gradient and is
comparable with that of the HPD in aerogel. Aerogel is known to have
a very broad fractal distribution of the particle correlation
length. The intrinsic dissipation of the CQP in aerogel can be
related with random spatial fluctuations of pairing potential of the
Cooper pairs caused by the random nature of aerogel structure. The
dipole potential is proportional to square of the pair condensation
energy. When the steady-state CQP is excited by applying a
sufficiently large rf-field at a frequency $f=f_0+\Delta f$ shifted
by $\Delta f$ from the cw-NMR frequency $f_0$, $\Delta f$ is
compensated by the tipping-angle-dependent dipole torque shown by
Fig.~\ref{schema}. When the pairing potential fluctuates, the
tipping angle and thus $M_\perp$ at position ${\bf r}$ fluctuates
from an average value $\overline{M}_\perp$ in such way that
\begin{equation}
M_\perp({\bf r}) = \overline{M}_\perp + \delta M_\perp({\bf r}).
\label{M_perp_in_CQP}
\end{equation}
Since $\delta M_\perp$ comes from the fluctuation of the dipole
potential, we can assume that
\begin{equation}
\delta M_\perp ({\bf r}) = \delta m({\bf r}) \cdot M_\perp,
\label{delta_M_perp_in_CQP}
\end{equation}
where non-dimensional quantity of the fluctuation $\delta m({\bf
r})$ related with dipole potential fluctuation is introduced.  A
typical length scale of spin motion in NMR should be the dipole
coherence length $\xi_D$ and the change of dipole potential should
be averaged in the scale smaller than $\xi_D$.  Thus there exists a
gradient of $M_\perp({\bf r})$ in aerogel such as,
\begin{equation}
\frac{\partial M_\perp({\bf r})}{\partial x_i} \sim \frac{\delta
m}{\xi_D} M_\perp \hspace{1mm} (x_i=1, 2, 3).
\label{devirative_of_M_perp_in_CQP}
\end{equation}
Here $\delta m({\bf r})$ is replaced by an averaged value $\delta
m$,
 and we assume three directions of magnetization gradients almost equally contribute to the energy loss.
When the magnetization is not uniform, spin diffusion takes place
and the dissipation due to spin diffusion is given by,
\begin{equation}
\dot{E} \sim -\sum_{i=1}^3 \int dv \frac{D_\perp}{\chi_A}
\frac{\partial M_\perp}{\partial x_i}
              \frac{\partial M_\perp}{\partial x_i},
\label{Energy_Loss_by_Diffusion_in_CQP}
\end{equation}
where $\chi_A$ is the susceptibility of A-phase.
According to this model, the size of the fluctuated region with
$\delta m$ is $\xi_D$ and numbers of the region per unit volume $N
\sim (1/\xi_D)^3$, and then
Eq.~(\ref{Energy_Loss_by_Diffusion_in_CQP}) becomes
\begin{equation}
\dot{E} \sim - 3 D_\perp \left( \frac{\delta m}{\xi_D} \right)^2
\frac{M_\perp^2}{\chi_A} (\xi_D^3 \cdot N) V,
\label{Energy_Loss_by_Diffusion_in_CQP_2}
\end{equation}
%
%
where $V$ is the volume of the sample and $(\xi_D^3 N) \sim 1$.
Fitting the observed loss, which is proportional to $M_\perp^2$ in
Fig.~\ref{absorption}, we obtained
\begin{equation}
\dot{E} = - 0.35 [{\rm nW}] \left( \frac{M_\perp}{M_\perp^{\rm max}}
\right)^2, \label{Exp_data_of_Energy_Loss_in_CQP}
\end{equation}
where $M_\perp^{\rm max} = \chi_A H$.
Combining Eqs.~(\ref{Energy_Loss_by_Diffusion_in_CQP_2}) and
(\ref{Exp_data_of_Energy_Loss_in_CQP}), $H=290$ Gauss and $D_\perp =
4 \times 10^{-3}$ cm$^2$/s~\cite{Sauls}, we get,
\begin{equation}
\left( \frac{\delta m}{\xi_D} \right) \sim 40 [{\rm cm}^{-1}].
\label{result_of_fluctuation_in_CQP}
\end{equation}
For $\xi_D \sim 10 \mu$m, $\delta m \sim 0.04$. This value of
$\delta m \sim 0.04$ should be compared with the line width of
cw-NMR.  In our case, the line width is larger than that for $\delta
m$, and may be determined by the texture.
There are many reports on the phase diagram of A-like and B-like
phases, the transition from A-like to B-like phase shows
supercooling~\cite{DmitCO,HalpCO,OsCO}.
We reported in \cite{KunimatsuJETPLett} that the part of cw-NMR
spectrum with a larger dipole shift makes the transition from A-like
to B-like phase at higher temperatures.
The supercooling transition of our sample has a width of about $50
\mu$K~\cite{Yamashita}, that may be related with the fluctuation of
pairing condensation observed here in $\delta m$. $\delta m \sim
\delta \Delta / \Delta$, which should be compared with the width of
A-B transition upon cooling $\Delta T_{AB} / T_c \sim 0.05$ mK /
2.03 mK $\sim 0.025$.

Similarly dissipation in the HPD can be calculated by the same
model. In the case of the HPD, the field gradient $G$ is applied to
excite the well-defined and stable HPD, and $\delta M_\perp({\bf
r})$ is give by,
\begin{equation}
\delta M_\perp({\bf r}) \sim \delta m({\bf r}) \left( \frac{\gamma G
\cdot z}{\Omega_B^2/\omega_L} \right) M_\perp(104^\circ),
\label{delta_M_perp_in_HPD}
\end{equation}
where $\Omega_B$ is the longtudinal angular frequency of superfluid
$^3$He B-phase, $\gamma$ is the gyromagnetic ratio and
$M_\perp(104^\circ) = \chi_B H \sin 104^\circ$. Substituting Eq.(10)
into Eq.(6) and replacing $\chi_A$ by $\chi_B$, the energy loss in
the HPD  is given in terms of $M_\perp$ by,
\begin{eqnarray}
\dot{E} &=& - \frac{\mu_0 D_\perp}{\chi_B} \left( \frac{\delta m}{\xi_D} \right)^2 \nonumber \\
     && \times \left( \frac{\gamma G L}{\Omega_B^2/\omega_L} \right)^2
        V \frac{M_\perp^3}{M_\perp^{max}}.
\label{Energy_Loss_in_HPD_2}
\end{eqnarray}
As shown in Fig.~\ref{absorption} where the field gradient of $G =
2.8$ mT/m was applied for the HDP data, the loss is not proportional
to $M_\perp^3$ and is much bigger than that given by
Eq.~(\ref{Energy_Loss_in_HPD_2}) for reasonable parameters chosen.
It is known that the large dissipation in the HPD comes from the
boundary layer of the domain, which may be the main contribution for
this thin sample.

In conclusion, the CQP in A-like phase in aerogel was first
observed. The CQP are stabilized by the orientation effect of the
global anisotropy in aerogel.  The stability of the CQP indicates
that the macroscopic phase coherence of precessing magnetization is
established for the whole sample of A-like phase in aerogel and the
long-range of the $\hat{l}$-vector is restored in aerogel.  The
dissipation of the CQP is caused by fluctuation of the pairing
potential averaged over the dipole coherence and our result of the
size of fluctuation may be consistent with the width of the
supercooling transition from A-like to B-like phase in aerogel.

We are thankful to
V.~V.~Dmitriev, M.~Krusius and G. E. Volovik for stimulating
discussions. The experimental part of this work was performed at
ISSP and was supported by the Joint CNRS-JSPS Project PRC-88, by the
21st Century COE Program and by KAKENHI program (grant 17071009).
Yu.~M.~B. is thankful for ISSP's visiting Professorship.


\end{document}